\documentclass[a4paper]{jpconf}
\usepackage{graphicx}
\usepackage{color}
\usepackage{ulem}
\usepackage{iopams}

\newcommand{\chiSG}{\ensuremath{\chi_\mathrm{SG}}}
\newcommand{\TSG}{\ensuremath{T_\mathrm{f}}}
\newcommand{\Tc}{\ensuremath{T_\mathrm{c}}}
\newcommand{\Ns}{\ensuremath{N_\mathrm{s}}}
\newcommand{\qEA}{\ensuremath{q_\mathrm{EA}^2}}

\begin{document}
\title{Critical property of spin-glass transition in a bond-disordered classical antiferromagnetic Heisenberg model with a biquadratic interaction}

\author{Hiroshi Shinaoka$^{1}$, Yusuke Tomita$^2$ and Yukitoshi Motome$^3$}
\address{$^1$Nanosystem Research Institute, AIST, Tsukuba 305-8568, Japan}
\address{$^2$Institute for Solid State Physics, University of Tokyo, Kashiwanoha, Kashiwa, Chiba, 277-8581, Japan}
\address{$^3$Department of Applied Physics, University of Tokyo, 7-3-1 Hongo, Bunkyo-ku, Tokyo 113-8656, Japan}

\ead{h.shinaoka@aist.go.jp}
\begin{abstract}
	Motivated by puzzling spin-glass behaviors observed in many pyrochlore-based magnets, effects of magnetoelastic coupling to local lattice distortions were recently studied by the authors for a bond-disordered antiferromagnet on a pyrochlore lattice [Phys. Rev. Lett. {\bf 107}, 047204 (2011)].
Here, we extend the analyses with focusing on the critical property of the spin-glass transition which occurs concomitantly with a nematic transition. 
Finite-size scaling analyses are performed up to a larger system size with 8192 spins to estimate the transition temperature and critical exponents. 
The exponents are compared with those in the absence of the magnetoelastic coupling and with those for the canonical spin-glass systems. 
We also discuss the temperature dependence of the specific heat in comparison with that in canonical spin-glass systems as well as an experimental result.
\end{abstract}

\section{Introduction}
Spin glass (SG), in which spins are frozen in random directions, is a low-temperature($T$) phase widely observed in geometrically frustrated magnets.
Recently, various puzzling behaviors have been observed experimentally for the SG transition in several pyrochlore-based magnets.
One is the robustness of the SG transition temperature $\TSG$.
For example, pyrochlore oxides $R_2$Mo$_2$O$_7$ show almost constant $\TSG \simeq 20$K 
for $R$=Y, Dy, and Tb despite largely different Curie-Weiss temperatures~\cite{Sato87,Ali89,Katsufuji00,Hanasaki06,Iguchi09}.
Furthermore, in
Y$_2$Mo$_2$O$_7$, $\TSG$ is practically unchanged against random substitution of Y$^{3+}$ by La$^{3+}$~\cite{Greedan86}.
Similar behaviors are observed in another pyrochlore-based magnets, e.g., spinel oxides $R$Cr$_2$O$_4$. 
In ZnCr$_2$O$_4$, random substitution of Zn$^{2+}$ by Cd$^{2+}$ destroys a spin-lattice ordering and 
induces a SG state with fairly flat $\TSG$~\cite{Ratcliff02}.
Another puzzling aspect is that $\TSG$ is much higher than a theoretically expected value for a moderate strength of disorder~\cite{Saunders07, Andreanov10,Tam10}.
These behaviors suggest that $\TSG$ is determined neither by a typical energy scale of exchange interactions nor the strength of disorder; 
some important factor is missing in the previous SG theories to understand the SG behaviors in these frustrated magnets.

Stimulated by these puzzles, 
the authors recently developed a SG theory by including a magnetoelastic coupling to local lattice distortions in a bond-disordered pyrochlore Heisenberg antiferromagnet~\cite{Shinaoka10b}. 
Using classical Monte Carlo (MC) simulations, we revealed that $T_\mathrm{f}$ is largely enhanced by the spin-lattice coupling.
Moreover, we showed that $T_\mathrm{f}$ is set by the energy scale of the spin-lattice coupling, and becomes almost independent of the disorder strength of the exchange interaction in a wide range of parameters.
These results explain well the puzzling aspects of the SG transition observed in the pyrochlore-based magnets.

An interesting question is whether the critical property of the SG transition is modified by the magnetoelastic coupling.
In the previous study, using finite-size scaling analyses, we showed that the MC data are compatible with a second-order transition,
and obtained the critical exponents $\gamma=1.2(5)$ and $\nu=0.67(16)$~\cite{Shinaoka10b}.
These critical exponents agree with the previously-reported ones in the absence of the spin-lattice coupling within the error bars~\cite{Saunders07,Andreanov10,Tam10}. 
The values are also consistent with recent numerical estimates for the canonical SG~\cite{Nakamura02,Nakamura03}.
These results suggest that the magnetoelastic coupling as well as the geometrical frustration does not modify the critical property.
The analyses, however, were limited to small system sizes, i.e., up to 2000 spins in Ref.~\cite{Shinaoka10b}.
It is well known that a careful consideration on finite-size effects is necessary in estimating the critical exponents for the canonical SG.
Here, we report more elaborate analysis up to a larger size with 8192 spins, and discuss the critical property in more detail. 

\section{Model and method}
We consider the following antiferromagnetic Heisenberg model with a biquadratic interaction introduced in Ref.~\cite{Shinaoka10b}:
\begin{equation}
	\mathcal{H}= \sum_{\langle i,j \rangle} 
	\Big[ J_{ij} \vec{S}_i \cdot \vec{S}_j - b_{ij} 
	\big( \vec{S}_i \cdot \vec{S}_j 
	\big)^2 
	\Big], \label{eq:Ham}
\end{equation}
where $\vec{S}_i$ denotes a classical Heisenberg spin at site $i$ and the sum runs over the nearest neighbor bonds of the pyrochlore lattice.
The model is derived from a spin-Peierls type model by integrating out local lattice distortions. 
Here, $J_{ij}$ is a bond-disordered exchange interaction, randomly chosen from $[J-\Delta,J+\Delta]$ with $0\le \Delta < J$. 
Effects of the magnetoelastic coupling are taken into account by the biquadratic interaction $b_{ij}$; 
we measure its strength by $b~(\equiv b_{ij}/J_{ij}^2)$. 

The previous MC study showed that the model exhibits a peculiar multicritical phase diagram between a nematic phase and SG phase~\cite{Shinaoka10b}. 
In the weakly disordered region $\Delta \lesssim b$, there are successive two transitions, 
a nematic transition at $\Tc \simeq b$ and a SG transition at $\TSG < \Tc$. 
$\TSG$ grows as $\Delta$ increases, and appears to merge into $\Tc$ at $\Delta \sim b$.
For larger $\Delta$, $\TSG~(=\Tc)$ becomes nearly independent of $\Delta$. 
In the following, we focus on the critical property of the concomitant transition in the region of $\Delta \gtrsim b$.
Specifically, we take $b=0.2$ as in the previous work, and perform MC simulations at $\Delta=0.8$.

In the following calculations, 
we consider periodic systems of cubic geometry with $L^3$ unit cells with totally $\Ns = 16L^3$ spins.
The largest system employed in the following simulations is $L=8$ with $8192$ spins.
In the MC sampling, we adopt the exchange MC method~\cite{Hukushima96} and the overrelaxation update~\cite{Alonso96} in addition to the standard single-spin-flip update.
To clarify the critical behaviors of the SG and nematic transitions, 
we calculate the SG susceptibility $\chiSG$, nematic susceptibility $\chi_Q$, and specific heat $C$.
The spin-glass susceptibility $\chiSG$ is given by $\Ns \qEA$, where $\qEA$ is the Edwards-Anderson order parameter defined by $\qEA \equiv \Ns^{-1} \langle \langle \sum_{\mu,\nu=x,y,z} (\sum_{i=1}^{N_\mathrm{s}} S_{i\mu}^\alpha S_{i\nu}^\beta)^2 \rangle_T \rangle_\Delta$~\cite{Edwards75}. 
Here $\langle \cdots \rangle_T$ denotes a thermal average and $\langle \cdots \rangle_\Delta$ a random average over the interaction sets; 
$\alpha$ and $\beta$ denote two independent replicas of the system with the same interaction set.
The nematic susceptibility $\chi_Q$ is given by $\chi_Q\equiv\Ns Q^2/T$, where $Q^2$ is the spin collinearity defined by $Q^2 \equiv 2 \Ns^{-2} \langle \langle \sum_{ij} \{ (\vec{S}_i \cdot \vec{S}_j)^2 - 1/3 \} \rangle_T \rangle_\Delta$.
The specific heat $C$ is calculated by $C=\Ns^{-1} T^{-1} \langle \langle \mathcal{H}^2 \rangle_T - \langle \mathcal{H} \rangle_T^2 \rangle_\Delta$.
In the following MC simulations, we interpolate the calculated thermodynamic quantities using the histogram reweighting technique~\cite{Ferrenberg88}.
All the data shown below are averaged over a number of interaction sets varying from 256 to 2000 depending on $L$. 

\section{Results}
Figures~1 shows the MC results at $b=0.2$ and $\Delta=0.8$.
To examine the critical property of the concomitant transition of the nematic and SG,
we perform finite-size scaling analyses for $\chiSG$ and $\chi_Q$ separately.
Here we assume $\chi = L^{\gamma/\nu} f(L^{1/\nu}t)$, where $t=(T-T_\mathrm{t})/T_\mathrm{t}$ and ($\chi$, $T_\mathrm{t}$) is either ($\chiSG$, $\TSG$) or ($\chi_Q$, $\Tc$).
$\nu$ and $\gamma$ are the critical exponents for the correlation length and $\chi$, respectively.
As demonstrated in Figs.~2(a) and 2(b), we successfully obtained scaling collapses of $\chiSG$ and $\chi_Q$ for $4\le L\le 8$: 
From the scaling of $\chiSG$ [Figs.~2(a)], 
we obtained $\TSG=0.248(2)$, $\gamma=1.5(1)$, and $\nu=0.80(2)$, 
and from that of $\chi_Q$ [Figs.~2(b)],
$\Tc=0.249(1)$, $\gamma=1.6(1)$, and $\nu=0.78(2)$.
The results are summarized in Table~1 with the estimates by using a set of data for smaller system sizes $2\le L\le 4$. 
Note that we observed no significant system-size dependence in the scaling results:
The estimates for different range of $L$ coincide with each other within the error bars. 
The results that the two transitions have the identical exponents and transition temperature confirm the conclusions in the previous study~\cite{Shinaoka10b}:
Two transitions occur concomitantly in a second-order fashion, and the critical exponents are indistinguishable from those for $b=0$~\cite{Saunders07,Andreanov10,Tam10} and the canonical SG~\cite{Nakamura02,Nakamura03}.

Figure~1(c) shows the $T$ dependence of $C$ for various system sizes.
$C$ exhibits a single broad peak at $T\simeq \TSG(=\Tc)$, in contrast to the first-order nature of the nematic transition at $\Delta=0$. 
The peak value $C_\mathrm{peak}$ shows a very weak $L$ dependence, and its growth gets slower as $L$ increases, as shown in the inset of Fig.~1(c).
When we fit the data by $C_\mathrm{peak}\propto L^\alpha$, 
the exponent $\alpha$ decreases as $L$ increases [$\alpha=0.076(1)$ for $2\le L\le 4$, and $\alpha=0.046(2)$  for $5\le L\le 8$].
Alternatively, the data can be well fitted by assuming $C_\mathrm{peak}^{-1}=aL^p+C_\mathrm{peak}^{-1}(\infty)$ with $p \simeq -1$ (not shown). 
These results suggest that $C$ is non-singular in the thermodynamic limit.
The broad peak behavior is superficially similar to that observed in the canonical SG~\cite{Binder86}, but
the peak is located near $\TSG$ in the present case: the peak position is estimated as $T_\mathrm{peak}=0.248(5)$ for $L=8$.
This is in contrast to the case of the canonical SG in which $T_\mathrm{peak}$ exceeds $\TSG$ typically by 20\%~\cite{Binder86}.
The broad peak at $\TSG$ will be of characteristic of the SG transition concomitant with the nematic transition in the present system.
Note that the result is consistent with the magnetic specific heat measured for Y$_2$Mo$_2$O$_7$, in which a broad peak is observed near $\TSG\simeq 20$K~\cite{Raju92}.
\begin{figure}
 \centering
  \resizebox{1.0\textwidth}{!}{\includegraphics{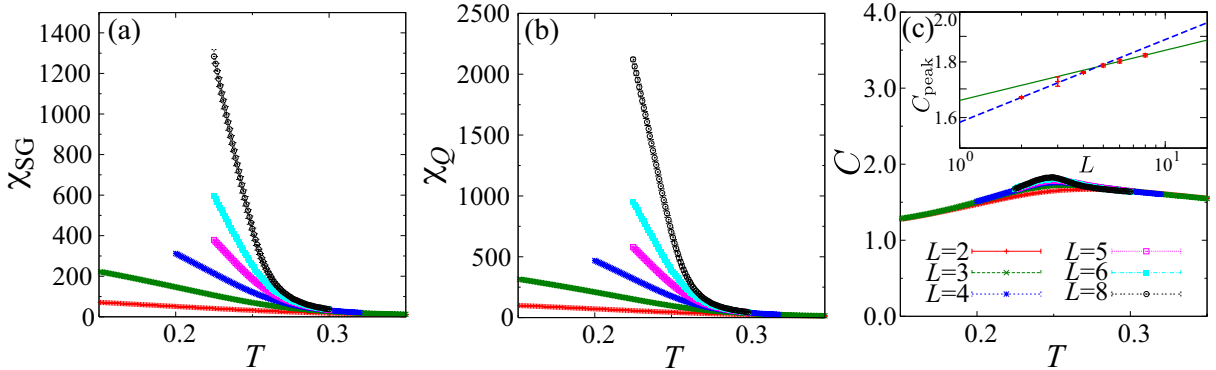}}
 \caption{\label{Fig1} Calculated thermodynamic quantities at $b=0.2$ and $\Delta=0.8$:
(a) the spin-glass susceptibility $\chiSG$, (b) the nematic susceptibility $\chi_Q$, and (c) the specific heat $C$.
The system-size dependence of the peak value of $C$ is shown by a log-log plot in the inset of (c).
}
 \label{fig:1}
\end{figure}
\begin{figure}
 \centering
 \includegraphics[width=.7\textwidth,clip]{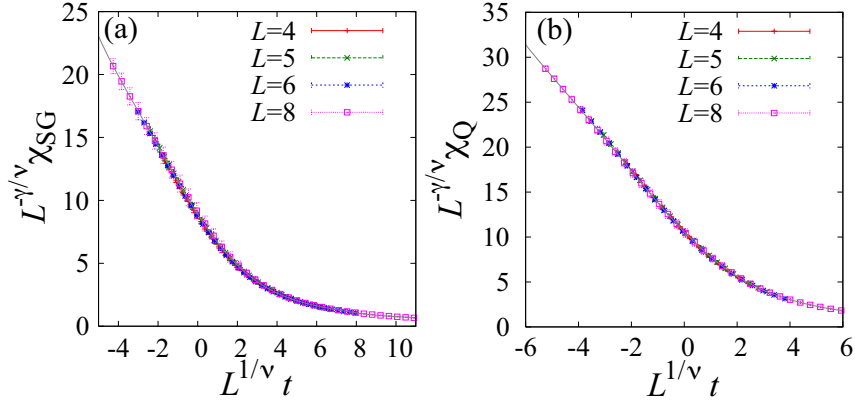}\hspace{2pc}
 \begin{minipage}[b]{8pc}
	 \caption{\label{Fig2} Scaling collapses of (a) $\chiSG$ and (b) $\chi_Q$ for the data in the range of $0.225\le T\le 0.275$.}
 \end{minipage}
 \label{fig:1}
\end{figure}
\begin{table}
 \centering
 \begin{tabular}{l|lll|lll}\hline
	 System size & \multicolumn{3}{c|}{$\chiSG$}      & \multicolumn{3}{c}{$\chi_Q$}  \\ \cline{2-7}
               & $\TSG$       & $\gamma$ & $\nu$   & $\Tc$    & $\gamma$ & $\nu$   \\ \hline
   $L=2,3,4$   &  0.248(2)    & 1.57(7)  & 0.79(2) & 0.242(2) & 2.0(3)   & 0.87(7) \\
   $L=4,5,6,8$ &  0.248(2)    & 1.5(1)   & 0.80(2) & 0.249(1) & 1.6(1)   & 0.78(2) \\ \hline
 \end{tabular}
 \caption{Estimates of the transition temperatures and critical exponents. The system sizes employed in the finite-size scaling analyses are $L=2,3,4$ and $L=4,5,6,8$, respectively.}
 \label{table:param}
\end{table}

\section{Summary}
In this paper, we have examined the critical property of a concomitant spin-glass and nematic transition in a bond-disordered pyrochlore antiferromagnet with a biquadratic interaction, which was recently found by the authors~\cite{Shinaoka10b}.
We have extended Monte Carlo simulations and finite-size scaling analyses up to a larger system size with 8192 spins.
As a result, we have confirmed the previous report that the concomitant transition is second order, and its critical property does not contradict those in the absence of the spin-lattice coupling and for the canonical spin glass.
Moreover, we have discussed the temperature dependence of the specific heat.

We thank H. Kawamura, H. Takayama, and T. Taniguchi for fruitful discussions.
This work was supported by Grant-in-Aids (No. 19052008), Global COE Program ``the Physical Sciences Frontier'', and HPCI Strategic Program, from MEXT, Japan.

\end{document}